\documentclass[preprint,showpacs,preprintnumbers,amsmath,amssymb,12pt]{article}

\usepackage{graphicx}
\usepackage{color}
\usepackage{amsfonts}

\renewcommand{\theequation}{\arabic{section}.\arabic{equation}}

\newcommand{\p}{\ensuremath{\phi}}
\newcommand{\s}{\ensuremath{\sigma}}

\renewcommand{\t}{\ensuremath{\tau}}

\def\gappeq{\mathrel{\rlap {\raise.5ex\hbox{$>$}}
{\lower.5ex\hbox{$\sim$}}}}

\def\lappeq{\mathrel{\rlap{\raise.5ex\hbox{$<$}}
{\lower.5ex\hbox{$\sim$}}}}

\def\ZM{{\mathbb Z}}

\def\CM{{\mathbb C}}

\def\a{\alpha}

\def\e{\epsilon}

\def\k{\kappa}
\def\l{\lambda}
\def\m{\mu}
\def\n{\nu}

\def\p{\pi}

\def\r{\rho}
\def\s{\sigma}
\def\t{\tau}

\def\F{\Phi}

\def\P{\Pi}

\def\ZM{{\mathbb Z}}

\newcommand{\be}{\begin{equation}}
\newcommand{\ee}{\end{equation}}

\newcommand{\ba}{\begin{eqnarray}}
\newcommand{\ea}{\end{eqnarray}}

\newcommand{\ns}{\normalsize}

\begin{document}


\begin{titlepage}


\title{
   \hfill{\ns CERN-PH-TH/2004-189\\}
   \vskip 2cm
   {\Large\bf Non-perturbative vacua for M-theory on $G_2$ manifolds}\\[0.5cm]}
\setcounter{footnote}{0}
\author{
{\ns Beatriz de Carlos$^1$\footnote{email:
B.de-Carlos@sussex.ac.uk \newline Also at: Department of Physics CERN, Theory
Division, 1211 Geneva 23, Switzerland}}~,
\setcounter{footnote}{3}
{\ns Andr\'e Lukas$^1$\footnote{email: A.Lukas@sussex.ac.uk}}~
  {\ns and Stephen Morris$^2$\footnote{email: smorris@perimeterinstitute.ca}}
  \\[0.5cm]
   {\it\ns $^1$Department of Physics and Astronomy, University of Sussex}\\
   {\ns Falmer, Brighton BN1 9QJ, UK} \\[0.5cm]
   {\it\ns $^2$Perimeter Institute for Theoretical Physics}\\
   {\ns Waterloo, Ontario N2L 2Y5, Canada}}

\date{\today}

\maketitle

\begin{abstract}\noindent
We study moduli stabilization in the context of M-theory on
compact manifolds with $G_2$ holonomy, using superpotentials from
flux and membrane instantons, and recent results for the K\"ahler
potential of such models. The existence of minima with negative
cosmological constant, stabilizing all moduli, is established.
While most of these minima preserve supersymmetry, we also find
examples with broken supersymmetry. Supersymmetric vacua with
vanishing cosmological constant can also be obtained after a
suitable tuning of parameters.
\end{abstract}


\thispagestyle{empty}

\end{titlepage}


\section{Introduction}

The stabilization of moduli is a long-standing problem in string
theory.  Traditionally, attempts to solve it have included
non-perturbative effects, such as gaugino condensation, that generate
a (super)-potential for the moduli fields. There is a substantial body
of literature on this subject, particularly in the context of
heterotic string theory, see for
example~\cite{Ferrara:1982qs}--\cite{Lukas:1997rb}. 
It is fairly difficult to find models
with proper minima in this way, and it is probably fair to say that
successful models constructed along these lines usually require
special parameter choices and some degree of
tuning~\cite{Dixon:1990ds}. The
resulting minima are quite shallow and only separated from runaway
directions by a small barrier~\cite{Brustein:1992nk,Barreiro:1998aj}.
 These problems can be traced back to the
nature of non-perturbative superpotentials, which are
double-exponential in (canonically normalized) moduli fields.

Recently, progress has been made by using flux of anti-symmetric
tensor fields to stabilize moduli~\cite{Kachru:2003aw}. Unlike
non-perturbative superpotentials, superpotentials from flux are merely
single-exponential in canonically normalized moduli fields and are,
therefore, more likely to produce minima.  It has been known for some
time~\cite{Dine:1985rz,Rohm:1985jv,Wen:1985qj,Strominger:1986uh,CH}
that a combination of flux and non-perturbative effects can stabilize
moduli successfully, and under relatively generic conditions, but only
with recent advances in the understanding of flux
compactifications~\cite{GVW}--\cite{Lukas:2004ip} has this possibility
been analyzed more
systematically~\cite{Acharya:2002kv}--\cite{Gukov:2003cy}. In this
paper we will present the first systematic analysis of this problem
for M-theory on manifolds of $G_2$ holonomy with a superpotential from
flux and membrane instantons. 

\vspace{0.4cm}

M-theory on (compact) manifolds $X$ of $G_2$ holonomy leads to
non-chiral~\cite{Witten:1983ux} four-dimensional effective theories
with $N=1$ supersymmetry~\cite{Papadopoulos:1995da,Beasley:2002db}.
Non-Abelian gauge groups and matter fields, which may account for the
standard model particles, arise when the $G_2$ space develops
singularities~\cite{Acharya:1998pm}--\cite{Berglund:2002hw}. In this
paper we will not consider such possible matter field sectors, but
focus on the gravity/moduli part of the theory, which contains
$b^3(X)$ chiral moduli multiplets $\F^I$. Their real parts
parameterize the moduli space of the $G_2$ manifold, while their
imaginary parts corresponds to axions. We are interested in
stabilizing these moduli fields $\F^I$, combining the effects of flux
and membrane instantons. The flux superpotential has been computed in
Ref.~\cite{SG,Beasley:2002db}, while the result for the structure of
membrane instanton contributions to the superpotential can be found in
Ref.~\cite{Harvey:1999as}. The scalar potential of four-dimensional
$N=1$ supergravity also depends on the K\"ahler potential for which we
rely on the results of Ref.~\cite{Lukas:2003dn}.  We will be focusing
on compact $G_2$ manifolds constructed by blowing up the singularities
of $G_2$ orbifolds~\cite{joyce1,joyce2,joyceb,Acharya:1998pm}. For
such $G_2$ manifolds, the moduli naturally split into two classes,
$(\F^I)=(T^a,U^i)$, namely the ``bulk'' moduli $T^a$ associated with
the underlying torus and the ``blow-up'' moduli $U^i$ that arise from
blowing up the singularities. We will address the stabilization of
both types of moduli for realistic models.

A practical problem is that there are no simple compact $G_2$
manifolds with, say, $b^3(X)=1$ available. In particular, the
calculation of the K\"ahler potential in Ref.~\cite{Lukas:2003dn}
has been carried out for an example with $b^3(X)=43$. We will,
therefore, have to deal with a large number of moduli. This task
will be approached by starting with relatively simple, but
characteristic, toy models and then working our way up to include
the full complications of more realistic cases. After setting up
the structure of the models in the next section,
Section~\ref{bulk} discusses the stabilization of the bulk moduli
$T^a$, first for a simple universal case, and then including all
bulk moduli. Section~\ref{blow-up} includes the blow-up moduli
$U^i$, again starting with a simple universal toy model and then
moving on to include all moduli. Conclusions and further
directions are presented in Section~\ref{conclusion}.

\section{General structure of four-dimensional effective
theories}\label{setup}

In this section, we will review the structure of four-dimensional
$N=1$ theories, which originate from M-theory on (compact) $G_2$
manifolds. These four-dimensional models will form the basis for
the subsequent analysis in this paper. For mathematical facts on
manifolds with $G_2$ holonomy we refer to Ref.~\cite{joyceb}. Many
of the physics aspects of the following presentation can be found
in Refs.~\cite{Papadopoulos:1995da,Beasley:2002db}.

\vspace{0.4cm}

Before we specialize to the models considered in this paper, let
us collect a number of useful general facts about M-theory on
$G_2$ manifolds. A seven-dimensional manifold $X$ with holonomy
$G_2$ has a vanishing first Betti number, $b^1(X)=0$, and its
cohomology is, therefore, characterized by $b^2(X)$ and $b^3(X)$.
When compactifying M-theory on such a manifold, the resulting
four-dimensional low-energy theory has $N=1$ supersymmetry and
contains $b^2(X)$ Abelian vector multiplets and $b^3(X)$ uncharged
chiral multiplets~\cite{Witten:1983ux,Papadopoulos:1995da}. We denote these latter
chiral fields, which are the primary focus of this paper, by
$\F^I$, $I,J,\dots =1,\dots ,b^3(X)$.

To compute the scalar potential we will need an explicit
expression for the K\"ahler potential of these chiral superfields.
Ideally, one would like a generic and explicit formulae that may
depend on topological data of the $G_2$ manifold in question, but
still applies to all $G_2$ manifolds. Unfortunately, no such
formula is known. We will go around this problem by focusing on a
particular class of $G_2$ manifold for which an (approximate) form
of the K\"ahler potential is known. However, let us first collect
some general facts. It is known~\cite{Beasley:2002db} that the
K\"ahler potential can be expressed in terms of the volume ${\cal
V}$ of the internal manifold as
\begin{equation}
 K = -3\ln\left(\frac{{\cal V}}{2\p^2}\right)\; .\label{KV}
\end{equation}
This relation implies that $K$ only depends on geometrical data
and is, hence, independent of the axionic fields, which we take to
be the imaginary parts of the chiral fields $\F^I$. We, therefore,
have
\begin{equation}
 K=K(\F^I+\bar{\F}^I)\; .
\end{equation}
Further, it can be shown~\cite{Beasley:2002db} that a rescaling of
the $G_2$ space by a factor $\l$, corresponding to a change
${\cal V}\rightarrow \l^7{\cal V}$ of the volume, leads to a scaling
$(\F^I+\bar{\F}^I)\rightarrow\l^3(\F^I+\bar{\F}^I)$. Hence,
the real parts of the chiral fields $\F^I$ measure three-dimensional
volumes. More precisely, they are proportional to the volumes of
(a basis of) three cycles of the $G_2$ manifold.

It can be shown by straightforward dimensional
reduction~\cite{Lukas:2003dn,Papadopoulos:1995da} that the gauge
kinetic functions $f$ of the gauge multiplets are of the form
\begin{equation}
 f = c_I\F^I\; ,
\end{equation}
where $c_I$ are constant coefficients that depend on the $G_2$
manifold and the gauge multiplet under consideration.

Perturbatively, and in the absence of flux, the superpotential $W$
vanishes. It has been shown~\cite{SG,Beasley:2002db}
that internal flux of the M-theory four-form field strength leads
to a superpotential
\begin{equation}
 W_{\rm flux} = m_I\F^I\; , \label{genflux}
\end{equation}
where the flux parameters $m_I$ are quantized. Non-perturbatively,
an additional contribution to the superpotential is generated by
membrane instantons~\cite{Harvey:1999as} wrapping three-cycles
within the $G_2$ manifold.  A typical instanton superpotential is
given by a sum of terms of the form
\begin{equation}
 W_{\rm inst}\sim \exp (-p_I\F^I)\; ,
\end{equation}
where the constants $p_I$ determine the three-cycle in question.
In this paper we will study the stabilization of $G_2$ moduli due
to a combination of flux and membrane instanton contributions to
the superpotential. Note that the flux terms~(\ref{genflux}) in
$W$ break the discrete shift symmetries of the axions that, even
after the inclusion of instanton effects, one would normally
expect. Hence field values that differ by integer multiples of
$2\p i$ are no longer equivalent, as would have been the case in
the absence of flux. In Ref.~\cite{Beasley:2002db}, these integer
shifts of the axions have been related to the possible values $h$
of the external part of the flux. More precisely, the external
flux is related to the axionic components of the superfields by
\begin{equation}
 h\sim{\cal V}^{-3}m_I\Im (\F^I)  \; ,\label{extflux}
\end{equation}
where we have dropped an overall constant.

\vspace{0.4cm}

We will focus on a particular class of compact $G_2$ manifolds, as
constructed by Joyce~\cite{joyce1,joyce2,joyceb}. These $G_2$
manifolds are built from seven dimensional $G_2$ orbifolds with
co-dimensional four singularities by blowing up their
singularities. Accordingly, the moduli $\F^I$ split into two
classes, namely the ``bulk moduli'' $T^a$, where $a,b,\dots =
1,\dots ,A$, associated with the underlying orbifold and the
``blow-up moduli'' $U^i$, where $i,j,\dots =1,\dots ,I$.
Geometrically, the real parts of the bulk moduli measure the radii
(or angles) of the underlying torus, while the blow-up moduli
measure the size and orientation of the blow-ups. It will be
convenient to split the moduli fields into their real and
imaginary parts by writing
\begin{equation}
 T^a=t^a+i\t^a\; ,\qquad U^i=u^i+i\n^i \; .
\end{equation}
Recall that, more precisely, the real parts $t^a$ and $u^i$ are proportional
to volumes of internal three cycles, while the fields
$\t^a$ and $\n^i$ are axions. We normalize the real parts so they measure
these three-cycle volumes in units of $T_2^{-1}$, the inverse membrane
tension.

Here we consider examples where the only moduli of the underlying
orbifold are its seven radii. Hence, we have seven bulk moduli,
each with a real part proportional to the volume of one of the
seven bulk three-cycles. The calculation of
Ref.~\cite{Lukas:2003dn} shows that the K\"ahler potential for
these seven bulk moduli is given by
\begin{equation}
 K_{\rm bulk} = -\sum_{a=1}^7\ln\left(T^a+\bar{T}^a\right)\; ,\label{Kbulk}
\end{equation}
as expected. The idea is to obtain an approximate K\"ahler
potential for all moduli by expanding in the typical size $\r$ of
the blow-ups. Blowing up co-dimension four singularities
introduces additional two-cycles, and the blow-up moduli $U^i$
should, therefore, scale with $\r^2$. The construction of
Ref.~\cite{joyce1,joyceb} also shows that the leading correction
to the orbifold volume due to the blow-ups arises at order $\r^4$.
The leading correction to Eq.~(\ref{Kbulk}) must, therefore, be
proportional to $(U^i+\bar{U}^i)^2$. The general K\"ahler
potential can then be written as~\cite{Lukas:2003rr}
\begin{equation}
 K = -\sum_{a=1}^7\ln\left(T^a+\bar{T}^a\right)
     +\sum_{i=1}^If_i(U^i+\bar{U}^i)^2\; ,\label{K}
\end{equation}
where $f_i$ are functions of the real parts $t^a$ of the bulk moduli.  Under a
rescaling of all moduli with $\l^3$ the first term in Eq.~(\ref{K})
already reproduces the correct scaling of $K$, or equivalently, via
Eq.~(\ref{KV}), of the volume ${\cal V}$.  Consequently, the second term in Eq.~(\ref{K})
should be invariant under the rescaling. This implies the functions
$f_i$ must be homogeneous of degree $-2$ in $t^a$. For
practical purposes, we will use the simple form
\begin{equation}
 f_i = 8\left[\P_{a=1}^7(T^a+\bar{T}^a)^{p_{ia}}\right]^{-1}\; ,
 \label{f}
\end{equation}
where the vectors ${\bf p}_{i}=(p_{ia})$ satisfy
\begin{equation}
 \sum_{a=1}^7p_{ia}=2\; , \qquad p_{ia}\in\{ 0,1\} \label{pcons}
\end{equation}
for all $i$. This form has indeed been found in the explicit
calculation of Ref.~\cite{Lukas:2003dn} and can be
expected to hold for a wider class of examples. We will use the
K\"ahler potential~(\ref{K}), with the functions $f_i$ given by
Eq.~(\ref{f}), as the basis of our analysis. We should stress that
this K\"ahler potential is approximate and only holds for
sufficiently small blow-ups. More precisely, one expects
corrections of order $u^4/t^4$ to Eq.~(\ref{K}) and one should,
therefore, require the ratio $u/t$ to be smaller than one. We will
discuss the constraints on the validity of Eq.~(\ref{K}) in more
detail below.

In this paper we keep the number $I$ of blow-up moduli arbitrary.
For the examples presented in Ref.~\cite{joyceb} it ranges from a
few to well over $100$, but we will see later that consistent
minima can sometimes only be obtained for sufficiently low $I$.
The structure of the powers $p_{ia}$ will not be of particular
importance for our analysis and, apart from the
constraints~(\ref{pcons}), we will keep them arbitrary. Despite
this level of generality, it will be useful for us to present a
concrete example for illustration. This example is based on the
orbifold $T^7/\ZM_2^3$, and contains twelve co-dimension four
singularities locally modeled on $T^3\times\CM^2/\ZM^2$. For the
details of the construction we refer to
Refs.~\cite{Lukas:2003dn,joyce1,joyceb}. The singularities are
removed by blowing-up the origin in $\CM^2/\ZM^2$. Ignoring some
subtleties associated with the smoothing required to join them to
the rest of the manifold (see \cite{Lukas:2003dn} for details),
the blow-ups support an Eguchi-Hanson metric

\begin{equation}
 ds_{\rm EH}^2 = \frac{dr^2}{1-\frac{\r^4}{r^4}}+r^2\left[\left(1-
                 \frac{\r^4}{r^4}\right)\s_1^2+\s_2^2+\s_3^2\right]\; ,
\end{equation}

\noindent where $\s^i$ are Maurer-Cartan one forms, and the
parameter $\r$ is the radius of the blow-up. Each blow-up is
associated with three moduli (giving $I=36$) whose real parts have
the form $\Re(U^i)\sim\r^2 (n^\a)^2$, where $\mathbf{n}$ is a unit
three-vector describing the blow-up's orientation. The real parts
of the seven bulk moduli, on the other hand, are products of three
of the radii of the underlying torus, chosen so that they are the
volumes of the three cycles preserved by the orbifolding. The $36$
blow-up moduli can be labeled by a triple of indices $(i)=(\t
,n,\a)$, where $\t = 1,2,3$, $n=1,2,3,4$ and $\a = 1,2,3$. The
index $\t$ refers to the three $\ZM_2$ symmetries of the orbifold
and it indicates that a given blow-up modulus is associated with a
fix point under the $\t^{\rm th}$ $\ZM_2$ symmetry. Further, $\a$
is an orientation index and $n$ numbers the moduli of a given
fixed point type and given orientation. The powers ${\bf p}_i$
only depend on the type $\t$ and the orientation $\a$ and the nine
relevant vectors ${\bf p}_{(\t ,\a )}$ are given in
Table~\ref{tab1}.
\begin{table}
 \begin{center}
 \begin{tabular}{|l|l|l|l|}
  \hline
  ${\bf p}_{(\t ,a)}$&$a=1$&$a=2$&$a=3$\\\hline
  $\t = 1$&$(1,0,0,0,0,1,0)$&$(0,1,0,0,1,0,0)$&$(0,0,1,1,0,0,0)$\\\hline
  $\t = 2$&$(1,0,0,0,0,0,1)$&$(0,0,1,0,1,0,0)$&$(0,1,0,1,0,0,0)$\\\hline
  $\t = 3$&$(1,0,0,1,0,0,0)$&$(0,0,1,0,0,1,0)$&$(0,1,0,0,0,0,1)$\\\hline
\end{tabular}
 \caption{Values of vectors ${\bf p}_{(\t ,a)}$ that define the
          moduli K\"ahler potential for the $G_2$ manifold based on
          $T^7/\ZM_2^3$.}
 \label{tab1}
 \end{center}
\end{table}

We will consider a superpotential $W$ that arises from a
combination of flux and membrane instantons. From the above
discussion this should have the form
\begin{equation}
 W =  \sum_{a=1}^7w(m_a,k_a,T^a)+\sum_{i=1}^Iw(\m_i,l_i,U^i)\; ,\label{W}
\end{equation}
where the function $w$ is defined by
\begin{equation}
 w(m,k,X) = mX+ke^{-X}\; .\label{w}
\end{equation}
The parameters $m_a$ and $\m_i$ are the flux parameters for the
bulk and blow-up moduli, respectively. They are quantized and we
will take them to be integers for simplicity. This can always be
arranged by absorbing the overall factor into the normalization of
the potential and rescaling $k_i$, $l_i$. Strictly, the
pre-factors $k_a$ and $l_i$ for the instanton contributions are
(non-exponential) functions of the moduli. However, these
functions have not been explicitly computed for the cases at hand
and we will take them to be constant and real, if only to make the
problem manageable.

\vspace{0.4cm}

To summarize, the models we are going to consider contain seven
bulk moduli $T^a$ and an arbitrary number $I$ of blow-up moduli,
with a K\"ahler potential~(\ref{K})--(\ref{pcons}) and a
superpotential~(\ref{W}), (\ref{w}). Let us now discuss the
constraint we should impose on moduli space. For the supergravity
approximation, which underlies the above expressions for $K$ and
$W$, to be valid we need all internal length scales to be larger
than the 11-dimensional Planck length. Given the normalization of
our moduli this means that we should require
\begin{equation}
 t^a\gappeq 1\; ,\qquad u^i\gappeq 1\; .\label{cons1}
\end{equation}
In addition, to suppress higher ${\cal O}(u_i^4)$ terms in the
K\"ahler potential, we should also ensure that
\begin{equation}
 4\sum_{i=1}^If_iu_i^2\ll 1\; .\label{cons2}
\end{equation}
Geometrically, this constraint implies that the fraction of the
orbifold volume ``taken away'' by the blow-ups is small or, in
other words, that we are not too far away from the orbifold limit.
While we clearly need to impose a constraint of this type it is,
in practice, difficult to decide how small the left-hand side of
Eq.~(\ref{cons2}) really has to be for the unknown corrections to
the K\"ahler potential to be negligible. We will later adopt a
more rigorous approach whereby we add hypothetical ${\cal
O}(u_i^4)$ terms to the K\"ahler potential and check whether they
leave the lowest order results essentially unchanged.  We will
then only accept results that pass this consistency test. Note
that Eq.~(\ref{cons1}) provides a lower bound on individual $u^i$
while the left-hand side of the constraint~(\ref{cons2}) scales
with their total number $I$. We, therefore, anticipate that
consistent results are more difficult to obtain as the number of
blow-up moduli increases.

\vspace{0.4cm}

In this paper we are interested in the scalar potential for the
fields $(\F^I)=(T^a,U^i)$, which is given by the standard $N=1$
supergravity expression
\begin{equation}
  V=e^K\left[ K^{I\bar{J}}F_I\bar{F}_{\bar{J}}-3|W|^2\right] \label{V}
\end{equation}
where $K_{I\bar{J}}$ is the K\"ahler metric
\begin{equation}
 K_{I\bar{J}} = \frac{\partial^2K}{\partial\F^I\partial\bar{\F}^{\bar{J}}}
\end{equation}
and $K^{I\bar{J}}$ its inverse. The F-terms $F_I$ are defined by
\begin{equation}
 F_I=W_I+K_IW\; ,
\end{equation}
where indices on $W$ or $K$ generally indicate derivatives,
that is,
\begin{equation}
 K_I=\frac{\partial K}{\partial\F^I}\; ,\qquad
 W_I=\frac{\partial W}{\partial\F^I}\; .
\end{equation}
Let us summarize the main general properties of this $N=1$
supergravity potential, relevant to our applications. A minimum of
the potential only respects supersymmetry if $F_I=0$ for all $I$.
Conversely, a solution to the F-equations
\begin{equation}
 F_I=0\label{F}
\end{equation}
constitutes a supersymmetric extremum of the potential. The second
derivatives of $V$ at a supersymmetric extremum are given by
\begin{eqnarray}
 V_{IJ}\left.\right|_{F_I=0} &=& -e^K\bar{W}F_{IJ}\label{V2}\\
 V_{I\bar{J}}\left.\right|_{F_I=0} &=& e^K\left[ K^{K\bar{L}}F_{KI}\bar{F}_{\bar{L}\bar{J}}
                                       -2K_{I\bar{J}}|W|^2\right]\; .\label{V2m}
\end{eqnarray}
In general each individual case has to be checked to see if this
matrix is positive definite and such an extremum is indeed a
minimum. Suppose that we have a family of potentials that depend
on a set of parameters and that a minimum has been found for a
particular point in parameter space. The fact that the Jacobi
matrix of $V$ is positive definite, and hence non-singular, at
this minimum, together with the implicit function theorem, then
tells us that minima exit in an open neighbourhood of this
particular point in parameter space. In some of our subsequent
examples, we will explicitly find minima for special non-generic
values of measure zero in parameter space. The above statement
then establishes the existence of minima for more generic
parameter choices in an open neighbourhood of parameter space.

The potential value at a supersymmetric
extremum is
\begin{equation}
 V_{\rm min} = -3e^K|W|^2\; ,\label{Vmin}
\end{equation}
and, hence, the cosmological constant
is either negative if ($W\neq 0$) or zero if ($W=0$), leading to
${\rm AdS}_4$ or four-dimensional Minkowski space, respectively.
Conversely, a positive cosmological constant implies broken
supersymmetry. Supersymmetric extrema with vanishing cosmological
constant are characterized by
\begin{equation}
 W_I=0\; ,\qquad W=0\; ,\label{susyzero}
\end{equation}
and are, hence, independent of the K\"ahler potential. They are
only minima if $W_{IJ}$ is non-singular, as can be seen from
Eqs.~(\ref{V2}), (\ref{V2m}) by setting $W=0$. Again, suppose we
have a family of superpotentials depending on a set of parameters.
Finding a solution to the equations~(\ref{susyzero}) will usually
imply a fine-tuning in parameters space, as is normally required
to set the cosmological constant to zero. Now consider the first
derivatives $F_{IJ}$ and $F_{I\bar{J}}$ at a solution
of~(\ref{susyzero}). They are given by $F_{IJ}=W_{IJ}$ and
$F_{I\bar{J}}=0$ and, hence, the block matrix consisting of
$F_{IJ}$, $F_{I\bar{J}}$ and their conjugates is non-singular.
From the implicit function theorem this establishes the existence
of supersymmetric minima (generically with negative cosmological
constants) in an open neighbourhood in parameter space, around the
special point where the cosmological constant vanishes.

\section{Stabilizing bulk moduli}
\label{bulk}

The structure of our models is fairly complicated and finding
minima of the potential in the general case would be a quixotic
task. The strategy we will adopt starts with simplified models,
gradually working our way up to include the full complexity of the
system. As we will see, the results for the simplified models can
be used to construct minima in the general case. In this section,
we focus on the bulk moduli $T^a$ and neglect the blow-up moduli
$U^i$. They will be included in the subsequent section.

\subsection{The universal case}
\label{singleT}

As a further simplification, we first consider a ``universal'' case
where we set all seven bulk moduli equal. Such a model then contains
a single superfield
\begin{equation}
 T=t+i\t\; .\label{Tu}
\end{equation}
with K\"ahler potential
\begin{equation}
 K=-7\ln (T+\bar{T})\label{Ku}
\end{equation}
and superpotential
\begin{equation}
 W=w(m,k,T)=mT+ke^{-T}\; .\label{Wu}
\end{equation}
We can think of $t$ as the breathing mode of the $G_2$ manifold.
Of course, we are not claiming the above model corresponds to a
particular $G_2$ manifold in any strict sense. It is merely a toy
model that incorporates the main features of the bulk sector of
our realistic models. In addition, we will see that the results
from this model can be transferred to realistic cases.

With this understood, we are now going to analyze the properties of the
above model. The F-term can be written as
\begin{equation}
 F_T = w'(m,k,T)-\frac{7}{2t}w(m,k,T)\; ,\label{FT}
\end{equation}
where $w'$ is the derivative of $w$ with respect to its last argument.
More explicitly, split up into real and imaginary part, this reads
\begin{eqnarray}
 \Re (F_T) &=& -\frac{m}{2}\left[5+\k\left(2+\frac{7}{t}\right)\cos(\t )
               e^{-t}\right] \label{ReFT}\\
 \Im (F_T) &=& \frac{m}{2}\left[\k\left(2+\frac{7}{t}\right)\sin(\t ) e^{-t}
               -\frac{7\t}{t}\right]\; ,\label{ImFT}
\end{eqnarray}
where we have defined
\begin{equation}
 \k = \frac{k}{m}\; .
\end{equation}
The scalar potential~(\ref{V}) for the K\"ahler potential~(\ref{Ku}) and a general
$W$ is given by
\begin{equation}
 V = \frac{C^2}{t^5}\left\{|W_T|^2-\frac{7}{t}\Re (W_T\bar{W})+\frac{7}{t^2}|W|^2\right\}\; ,
\end{equation}
while inserting the explicit form~(\ref{Wu}) of $W$ leads to
\begin{eqnarray}
 V &=& \frac{C^2}{t^5}\left\{ 1+\frac{7\t^2}{t^2}+\k\left[5\cos (\t )+(\cos (\t)
     -\t \sin (\t ) )\frac{7}{t}-\frac{14\t \sin (\t )}{t^2}\right] e^{-t}\right.\nonumber\\
   &&\left.\qquad +\k^2\left[ 1+\frac{7}{t}+\frac{7}{t^2}\right]e^{-2t}\right\}
     \; .\label{Vu}
\end{eqnarray}
Here the overall constant $C^2$ is given by
\begin{equation}
 C^2=\frac{m^2}{224}\; .
\end{equation}
It can be shown from the F-terms~(\ref{ReFT}), (\ref{ImFT}) that every
supersymmetric extremum is indeed a minimum of the potential. Conversely,
we have found numerically that every minimum of the potential~(\ref{Vu})
preserves supersymmetry. It is, therefore, sufficient to look at
the $F$-equations.

\vspace{0.4cm}

It is easy to show from Eqs.~(\ref{ReFT}) and (\ref{ImFT}) that
the solutions to the F-equations can be characterized as follows.
The values of $t$ at the minima are the solutions to the equation
\begin{equation}
 \cos f(t)=-\frac{1}{\k}\frac{5t}{2t+7}e^t\; ,\label{teq}
\end{equation}
where
\begin{equation}
 f(t) = \frac{1}{7}\sqrt{\k^2(2t+7)^2e^{-2t}-25t^2}\; ,\label{fdef}
\end{equation}
for which the signs of $\k$ and $\sin f(t)$ are the same. For each such
$t$, there are two associated solutions for the imaginary part $\t$,
namely
\begin{equation}
 \t = \pm f(t)\; .\label{tausol}
\end{equation}
From Eq.~(\ref{teq}) all (positive) solutions for $t$ must be in the range
\begin{equation}
 t\in [0,t_{\rm max}]\; ,
\end{equation}
where $t_{\rm max}$ is defined by
\begin{equation}
 \frac{5t_{\rm max}}{2t_{\rm max}+7}e^{t_{\rm max}}=|\k|\; .\label{tmax}
\end{equation}
Very roughly we have
\begin{equation}
 t_{\rm max}\sim \ln |\k|\; .\label{tmaxi}
\end{equation}
In order to have solutions that satisfy the
constraint~(\ref{cons1}) on $t$, we need the parameter $|\k|$ to
be sufficiently large, typically of order one or larger. We note
that while $t$ varies between zero and $t_{\rm max}$ the function
$f$ varies between $|\k|$ and zero. This means, from
Eq.~(\ref{teq}), that the typical number, $N_{\rm sol}$, of minima
is given by
\begin{equation}
 N_{\rm sol} \sim \frac{|\k|}{\p}\; , \label{Nsol}
\end{equation}
where we have remembered the two possibilities~(\ref{tausol}) for
the imaginary part $\t$. This relation should, of course, be
understood as a rough indication of the number of minima for
sufficiently large $|\k|$. It shows that, for $|\k|\gg 1$, their
number is substantial. For $\k<0$ we have a particular minimum
with vanishing imaginary part at
\begin{equation}
 t=t_{\rm max}\; ,\qquad \t = 0\; . \label{realmin}
\end{equation}
All other minima are complex and $t<t_{\rm max}$.
\begin{figure*}[!htb]
\begin{center}
\includegraphics[width=10cm]{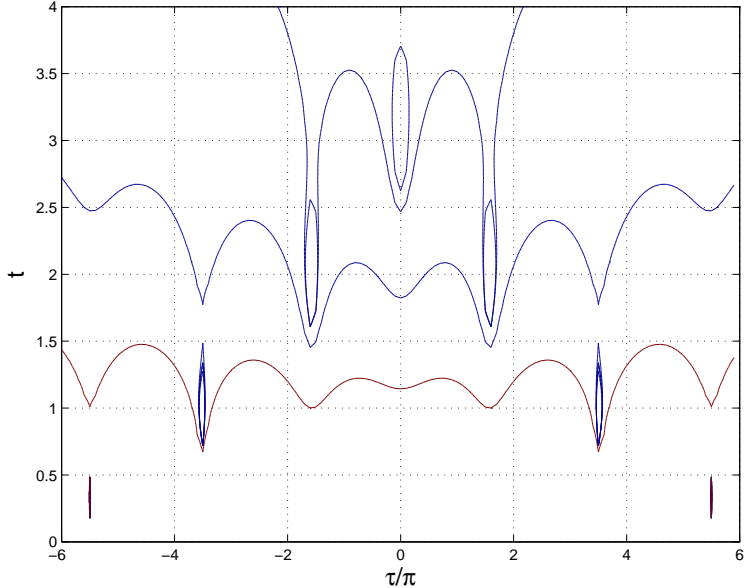}
\caption[fields]{\label{fig:singleT} Contour plot of the potential, given by Eq.~(\ref{Vu}),
in the ($t$,$\tau$) plane for $m=-1$, $k=20$.}
\end{center}
\end{figure*}
These statements can be easily checked in Figure~\ref{fig:singleT}, where we show a contour plot
of the scalar potential, as given by Eq.~(\ref{Vu}), in the $t$, $\tau$ plane. We have chosen
$m=-1$, $k=20$, and we can see that, following Eq.~(\ref{Nsol}), the number of minima is
very close to $|\kappa|/\pi$, with $\kappa=k/m=-20$; moreover, and according to Eq.~(\ref{tmaxi}),
the value of $t_{\rm max}$ is given by ${\rm ln} |\kappa|$ and, also, we can see in the
figure that Eq.~(\ref{realmin}) is fulfilled. Finally we have checked that all these minima
preserve supersymmetry with negative vacuum energy.

More precisely, from Eq.~(\ref{Vmin}) the potential value at any one of these minima is given by
\begin{equation}
 V_{\rm min} = -\frac{21C^2}{t^5}\frac{(t+1)^2+\t^2}{(2t+7)^2}\; ,
\end{equation}
and is, hence, always strictly negative. Also, since from
Eq.~(\ref{tausol}) $\t$ increases as $t$ decreases, the absolute
minimum will be the one with the smallest value of $t$. For
$\k>0$, or sufficiently negative $\k\lappeq -5$, this is always a
complex minimum, $\t\neq 0$. Also note that the two minima for a
given solution of $t$ are degenerate, since they only differ by a
sign of the imaginary part $\t$.

\vspace{0.4cm}

To summarize, for our simple model, the structure of minima is
controlled by the parameter $\k$, defined in Eq.~(\ref{K}). To
have minima compatible with the supergravity approximation, that
is, minima satisfying Eq.~(\ref{cons1}), we need $|\k|\gappeq 1$.
The number of minima is roughly given by $|\k|/\p$ and the real
part $t$ is always in the range $t\in [0,t_{\rm max}]$, where,
roughly, $t_{\rm max}\sim\ln |\k|$. For $\k<0$ a special real
minimum with $t=t_{\rm max}$ and $\t =0$ exists; all other minima
are complex. All minima preserve supersymmetry and have a negative
cosmological constant. The absolute minimum is the one with the
lowest value of $t$, and generally has a non-vanishing imaginary
part, $\t\neq 0$.

\subsection{Non-universal bulk moduli}

We now move on to a more realistic model with seven individual moduli
\begin{equation}
 T^a=t_a+i\t_a\; ,
\end{equation}
where $a,b,\dots = 1,\dots ,7$. The real parts $t_a$ can be
interpreted as the seven radii of the torus that underlies the
construction of the $G_2$ manifold. The K\"ahler potential and
superpotential are now given by
\begin{eqnarray}
 K &=& -\sum_{a=1}^7\ln (T^a+\bar{T}^a) \\
 W &=& \sum_{a=1}^7w_{(a)}\; ,
\end{eqnarray}
where the function $w$ is defined in Eq.~(\ref{w}) and we have
written
\begin{equation}
 w_{(a)}=w(m_a,k_a,T^a)\; ,
\end{equation}
for ease of notation. The F-terms take the form
\begin{equation}
 F_a = w_{(a)}'-\frac{1}{2t_a}W\; ,\label{Fa}
\end{equation}
while the scalar potential can be written as
\begin{equation}
 V = \frac{1}{32\P_{b=1}^7t_b}\left\{\sum_a\left[ t_a^2|w_{(a)}'|^2-t_a
     \Re (w_{(a)}'\bar{W})\right]+|W|^2\right\}\; . \label{pot7}
\end{equation}
Here $w'$ denotes the derivative of $w$ with respect to its last argument.
Analyzing the general vacuum structure of this model would be rather cumbersome
given the large parameter space. We will not attempt to do this but instead
present a number of examples.

\vspace{0.4cm}

Consider a ``universal'' choice of parameters, independent of the index $a$, that is,
\begin{equation}
 m_a=m\; ,\qquad k_a=k\label{mkuni}
\end{equation}
for some real $m$, $k$ and all $a=1,\dots ,7$. Then the F-terms~(\ref{Fa})
can be written as
\begin{equation}
 F_a = w'(m,k,T^a)-\frac{1}{2t_a}\sum_{b=1}^7w(m,k,T^b)\; .\label{Fa1}
\end{equation}
Now take any of the supersymmetric minima $T=t+i\t$ found in the previous
subsection for the universal case. Comparison of Eqs.~(\ref{FT}) and (\ref{Fa1})
then shows that setting
\begin{equation}
 T^a=T\; ,
\end{equation}
for all $a=1,\dots ,7$ solves all seven equations $F_a=0$. Hence, every supersymmetric
minimum of the universal single $T$ model leads to a supersymmetric extremum for
the case with seven moduli $T_a$, provided the parameters have been chosen
universally, as in Eq.~(\ref{mkuni}). It can be shown from Eqs.~(\ref{V2}) and (\ref{V2m})
that these extrema are indeed minima if $t$ is sufficiently large, that is,
roughly $t\gappeq 1$. As in the universal case, the cosmological constant is always
negative. As discussed at the end of Section~\ref{setup}, these minima still exist
for a non-universal choice of parameters sufficiently close to the universal one.

\vspace{0.4cm}

Allowing seven $T$-moduli also opens up qualitatively new possibilities compared
to the universal case. Consider the conditions~(\ref{susyzero}) for supersymmetric
minima with vanishing cosmological constant. Applied to the present case they
lead to moduli values
\begin{equation}
 T^a = \ln\left|\frac{k_a}{m_a}\right|+\p in_a \label{Tmin}
\end{equation}
where $n_a$ is an integer which is even for positive $k_a/m_a$ and odd otherwise.
In addition, vanishing of the real and imaginary part of the superpotential implies that
\begin{eqnarray}
 \sum_{a=1}^7m_a\left( 1+\ln\left|\frac{k_a}{m_a}\right|\right) &=& 0\label{tuning}\\
 \sum_{a=1}^7m_an_a &=& 0\; .\label{deg}
\end{eqnarray}
The first of these conditions is the usual fine-tuning required to
make the cosmological constant vanish. It constrains parameters to
a particular hyper-surface in parameter space. With favourable
choices for the signs of $k_a/m_a$, the second condition will have
an infinite number of solutions. They correspond to an infinite
number of degenerate supersymmetric minima with vanishing
cosmological constant that differ by integer shifts in the axion
directions.
\begin{figure*}[!htb]
\begin{center}
\includegraphics[width=10cm]{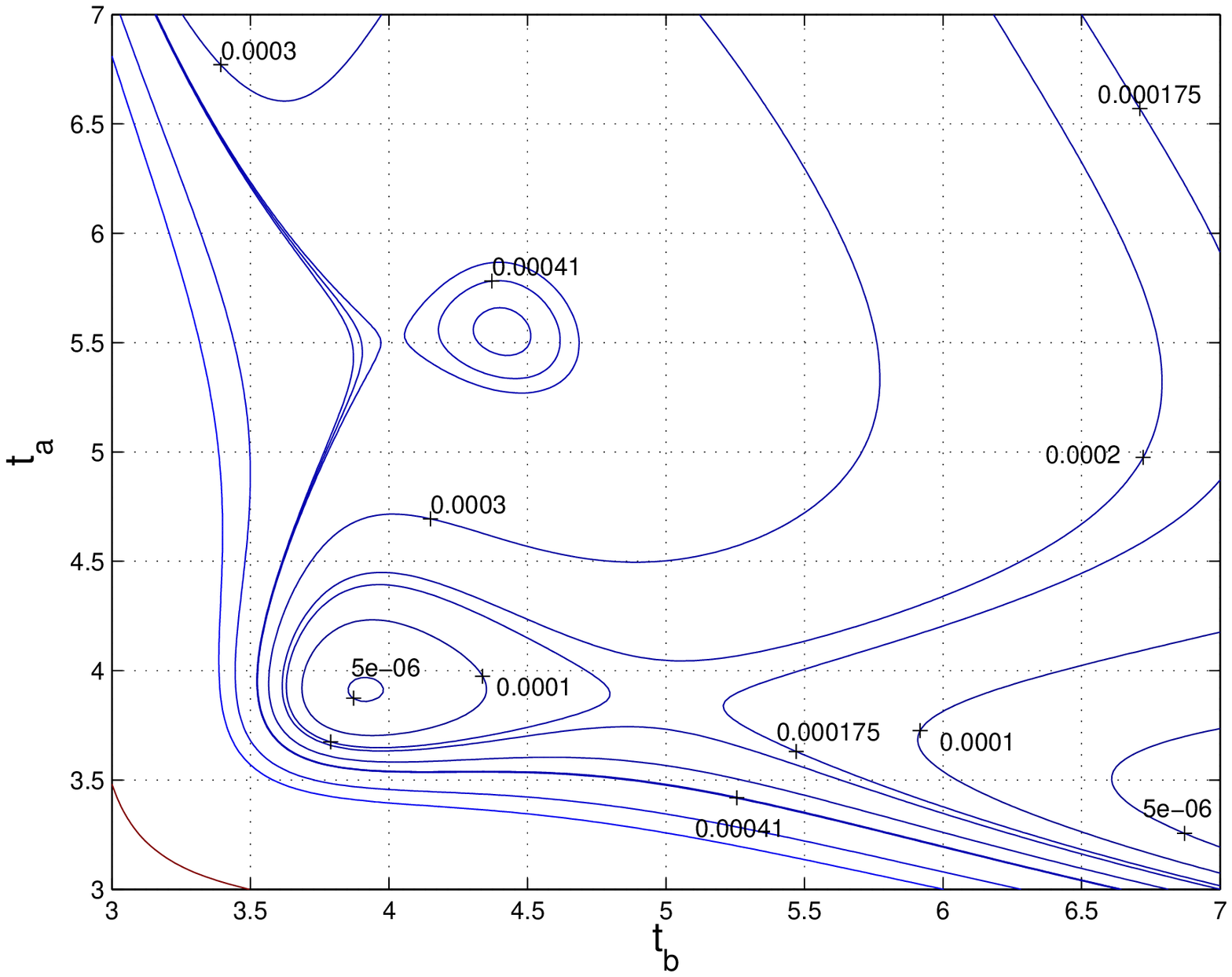}
\caption[fields]{\label{fig:cReTab} Contour plot of the potential,
given by Eq.~(\ref{pot7}), as a function of $t_a$ ($a=1,2,3$) and
$t_b$ ($b=4,5,6,7$) for $m_a=4$, $k_a=200$, $m_b=-3$, $k_b=-150$.
The imaginary parts of all fields have been set to zero, where we
have a minimum.}
\includegraphics[width=10cm]{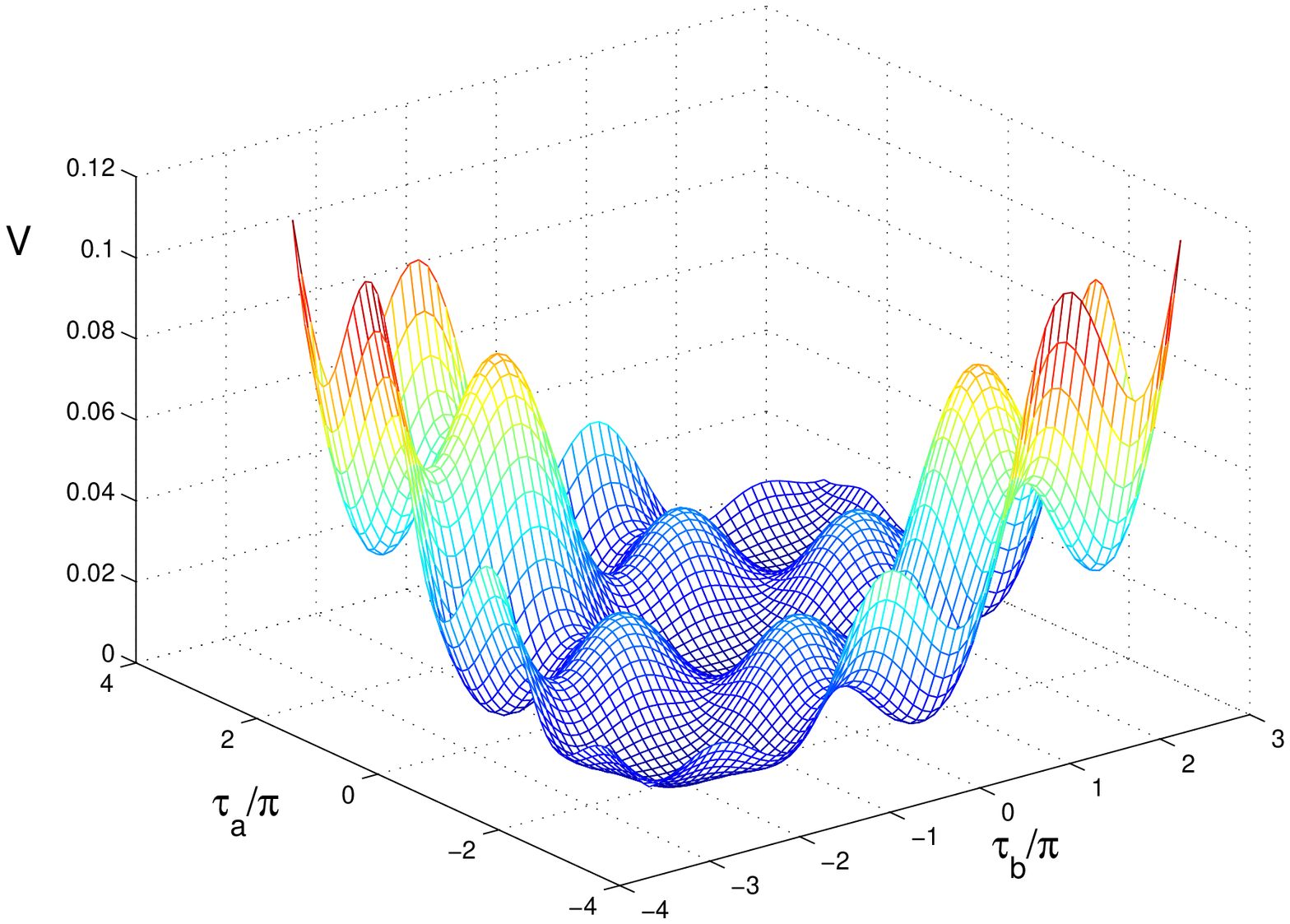}
\caption[fields]{\label{fig:ImTab} Plot of the potential, given by
Eq.~(\ref{pot7}), as a function of $\tau_a$, $\tau_b$ for the same
values of parameters as in the previous figure. The real parts of
the fields have been set to their minimum values at ${\rm ln}
|k_i/m_i|$, $i=a,b$.}
\end{center}
\end{figure*}
We present an example of this behaviour in
Figures~\ref{fig:cReTab},\ref{fig:ImTab}, where we have split the
moduli into two types, $a=1,2,3$ and $b=4,5,6,7$, in order to make
the graphics manageable. In Figure~\ref{fig:cReTab} we show a
contour plot of the potential, as given by Eq.~(\ref{pot7}), in
the $t_a$, $t_b$ plane, for values of the parameters $m_a=4$,
$k_a=200$, $m_b=-3$, $k_b=-150$, with the imaginary parts of the
fields, $\tau_a$, $\tau_b$ fixed at zero. In this way, both
Eqs~(\ref{tuning},\ref{deg}) are fulfilled. Moreover, a minimum is
found at the predicted values for the moduli, given by
Eq.~(\ref{Tmin}) with $n_a=n_b=0$, which, in this case, becomes
$t_a=t_b=3.9$. We can also see that for larger values of $t_a$ a
maximum appears, while for larger $t_b$ we have a saddle point. In
Figure~\ref{fig:ImTab}, where we show the potential as a function
of $\tau_a$, $\tau_b$ for $t_a$, $t_b$ fixed at their minimum
values, we can check how $\tau_a=\tau_b=0$ does indeed correspond
to a minimum which, however, is degenerate with those at
$\tau_a=\tau_b=2n\pi$ for $n$ integer. There exists, therefore, in
these cases, an array of degenerate minima with zero cosmological
constant. An obvious consequence of this pattern should be the
formation of domain walls between the different minima.

Comparison of Eqs~(\ref{extflux}) and (\ref{deg}) show that all
these minima correspond to vanishing external flux.\footnote{We
remark that minima with non-vanishing external flux can be
obtained for complex parameters $k_a$.}  From our general
discussion, we also know that supersymmetric minima, although with
negative cosmological constant, exist in a neighbourhood of the
hypersurface~(\ref{tuning}) in parameter space.

\section{Including blow-up moduli}
\label{blow-up}

We should now add blow-up moduli to the models of the previous
section. We begin with the universal case.

\subsection{The universal case}

An obvious extension of the simple model with a single bulk
modulus $T$ presented in Section~\ref{singleT} adds a second field
$U$ as the typical representative of a blow-up modulus. Split into
real and imaginary parts we write
\begin{eqnarray}
 T&=&t+i\t \\
 U&=&u+i\n \; .
\end{eqnarray}
The model is then defined by
\begin{eqnarray}
 K&=&-7\ln (T+\bar{T})+b\frac{(U+\bar{U})^2}{(T+\bar{T})^2} \label{KTU}\\
 W&=&w(m,k,T)+w(\m ,l,U)\; ,\label{WTU}
\end{eqnarray}
where $b$ is a real positive constant. To make contact with the
full model later on we will need to set $b=8I$, where $I$ is the
number of blow-up moduli. However, for the purpose of this
sub-section, we will treat $b$ as a phenomenological parameter. It
is useful to introduce the ratio
\begin{equation}
 \epsilon = \frac{u}{t}\; ,
\end{equation}
of blow-up and bulk modulus. Recall that the above K\"ahler
potential should be viewed as an expansion in $\e$ where terms of
order $\e^4$ and higher have been neglected. In accordance with
the general constraint~(\ref{cons2}) we should, therefore, work in
the region of moduli space where
\begin{equation}
 b\e^2 \ll 1\; .\label{consTU}
\end{equation}
Note that corrections of ${\cal O}(\e^4)$ to $K$ will lead to
${\cal O}(\e^2)$ terms in the scalar potential. Hence, on the
basis of the K\"ahler potential~(\ref{KTU}), we can reliably
compute the scalar potential only up to terms of ${\cal O}(\e )$.
Subsequent formulae will be quoted to this order.

It is crucial to check that relevant features of the scalar
potential, such as minima arising at ${\cal O}(\e )$, are stable
under inclusion of terms of ${\cal O}(\e^2)$ and higher, so
that~(\ref{consTU}) is indeed satisfied to the necessary degree.
In practice, we will add a hypothetical correction
\begin{equation}
\delta K = d\frac{(U+\bar{U})^4}{(T+\bar{T})^4}
\end{equation}
to the K\"ahler potential~(\ref{KTU}), where $d$ is a real number,
and compute the scalar potential up to order $\e^2$ by including
this correction. We will then compare the potentials at ${\cal
O}(\e )$ and ${\cal O}(\e^2)$, and only accept minima if they
consistently arise at both orders and for a reasonable range of
$d$.

\vspace{0.4cm}

For the K\"ahler potential~(\ref{KTU}), and a general
superpotential, the F-terms, to order $\e$, are given by
\begin{eqnarray}
 F_T &=& W_T-\frac{7}{2t}W \label{FT1}\\
 F_U &=& W_U+\frac{b\e}{t}W\; .\label{FU1}
\end{eqnarray}
To this order, the scalar potential reads
\begin{equation}
  V=\frac{1}{112t^5}\left\{\frac{1}{2}|W_T|^2+\frac{7}{4b}|W_U|^2
    +2\e\Re (W_T\bar{W}_U)-\frac{7}{2t}\Re (W_T\bar{W})
    -\frac{7\e}{2t}\Re (W_U\bar{W})+\frac{7}{2t^2}|W|^2\right\}\; .\
 \label{VWu}
\end{equation}

\vspace{0.4cm}

The explicit expressions for the F-terms and the scalar potential
are fairly complicated and are given in Appendix~\ref{appA}.
Inspection of the general scalar potential~(\ref{V}) shows that
\begin{equation}
 \frac{\partial V}{\partial\t}(\t =0,\n =0 )=0\; ,\qquad
 \frac{\partial V}{\partial\n}(\t =0,\n =0 )=0 \; ,
\end{equation}
so the potential is extremized in the imaginary directions at $\t
= \n =0$. Here we will focus on this real case, that is, we will
set $\t = \n =0$. Equations~(\ref{RFT})--(\ref{IFU}) show that
then the imaginary parts of the F-terms are automatically zero,
that is, $\Im (F_T)=\Im (F_U)=0$, while the real parts simplify to
\begin{eqnarray}
 \Re (F_T) &=& -\frac{m}{2}\left[5+\k\left(2+\frac{7}{t}\right)e^{-t}
               +\frac{7}{t}(Mu+Le^{-u})\right] \label{FTr}\\
 \Re (F_U) &\simeq& m\left[M-Le^{-u}+\frac{b\e}{t}(t+\k e^{-t})\right]\; .\label{FUr}
\end{eqnarray}
Here we have normalized our parameters with respect to the flux
$m$ by defining
\begin{equation}
 \k = \frac{k}{m}\; ,\qquad L=\frac{l}{m}\; ,\qquad M=\frac{\m}{m}\; .
\end{equation}
For vanishing imaginary parts the potential~(\ref{VTU}) then
simplifies to
\begin{eqnarray}
 V &=& \frac{C^2}{t^5}\left\{ 1+\k\left[ 5+\frac{7}{t}\right] e^{-t}
       \pm L\left[\frac{3u}{t}+\frac{7}{t}\right] e^{-u}
       +K^2\left[ 1+\frac{7}{t}+\frac{7}{t^2}\right] e^{-2t}\right. \nonumber \\
    && \qquad +7L^2\left[\frac{1}{2b}+\frac{u}{t^2}+\frac{1}{t^2}\right] e^{-2u}
        \pm LK\left[\frac{4u}{t}+\frac{7}{t}+\frac{7u}{t^2}+\frac{14}{t^2}
         \right] e^{-t-u}\label{V0} \\
    && \qquad\left. +\frac{7M^2}{2b}+4M\frac{u}{t}+KM\left[\frac{3u}{t}+
       \frac{7u}{t^2}\right] e^{-t}+ LM\left[\frac{7u^2}{t^2}+
       \frac{7u}{t^2}-\frac{7}{b}\right]\right\}\; .\nonumber
\end{eqnarray}
After minimizing this potential in $t$ and $u$ we still, of
course, have to check whether the extrema in the imaginary
directions are indeed minima.

\vspace{0.4cm}

Let us start by looking for supersymmetric minima with vanishing
cosmological constant, that is, solutions to the equations
$W_T=0$, $W_U=0$ and $W=0$. The first two of these equations
immediately imply that
\begin{eqnarray}
 T &=& \ln\left|\k\right| +\p ip\label{Tval}\\
 U &=& \ln\left|\frac{L}{M}\right|+\p iq\; ,\label{Uval}
\end{eqnarray}
where $p$ ($q$) is an integer which is even for $\k >0$ ($L>0$)
and odd otherwise. Vanishing of the superpotential leads to the
two conditions
\begin{eqnarray}
 m\left(1+\ln\left|\k\right|\right)
 +\m\left(1+\ln\left|\frac{l}{\mu}\right|\right)&=&0\; , \label{tuningTU}\\
 mp+\m q &=& 0\; .\label{degTU}
\end{eqnarray}
For fixed (integer) flux $m$ and $\m$ the first of these equations
can always be solved for appropriate choices of the parameters $k$
and $l$. The second equation has an infinite number of other
solutions when the signs of $\k$ and $\mu /l$ are chosen
appropriately. In particular, for $\k>0$ and $\mu /l>0$, we can
take $p=q=0$ and obtain a real supersymmetric minimum with
vanishing cosmological constant.  Hence, by tuning in parameter
space, we can have supersymmetric minima with vanishing
cosmological constant, infinitely degenerate by integer shifts in
the axion directions. From Eqs.~(\ref{extflux}) and (\ref{degTU})
all those minima correspond to vanishing external flux. In the
neighbourhood in parameter space of each such minimum there will
be supersymmetric minima with negative cosmological constant. We
note that their existence is independent of the K\"ahler
potential. Hence, we do not need to require that $b\e^2\ll 1$,
although this could be easily arranged by choosing parameters.
However, unless $b\e^2\ll 1$, we are unable to explicitly write
down the scalar potential close to these minima.
\begin{figure*}[!htb]
\begin{center}
\includegraphics[width=10cm]{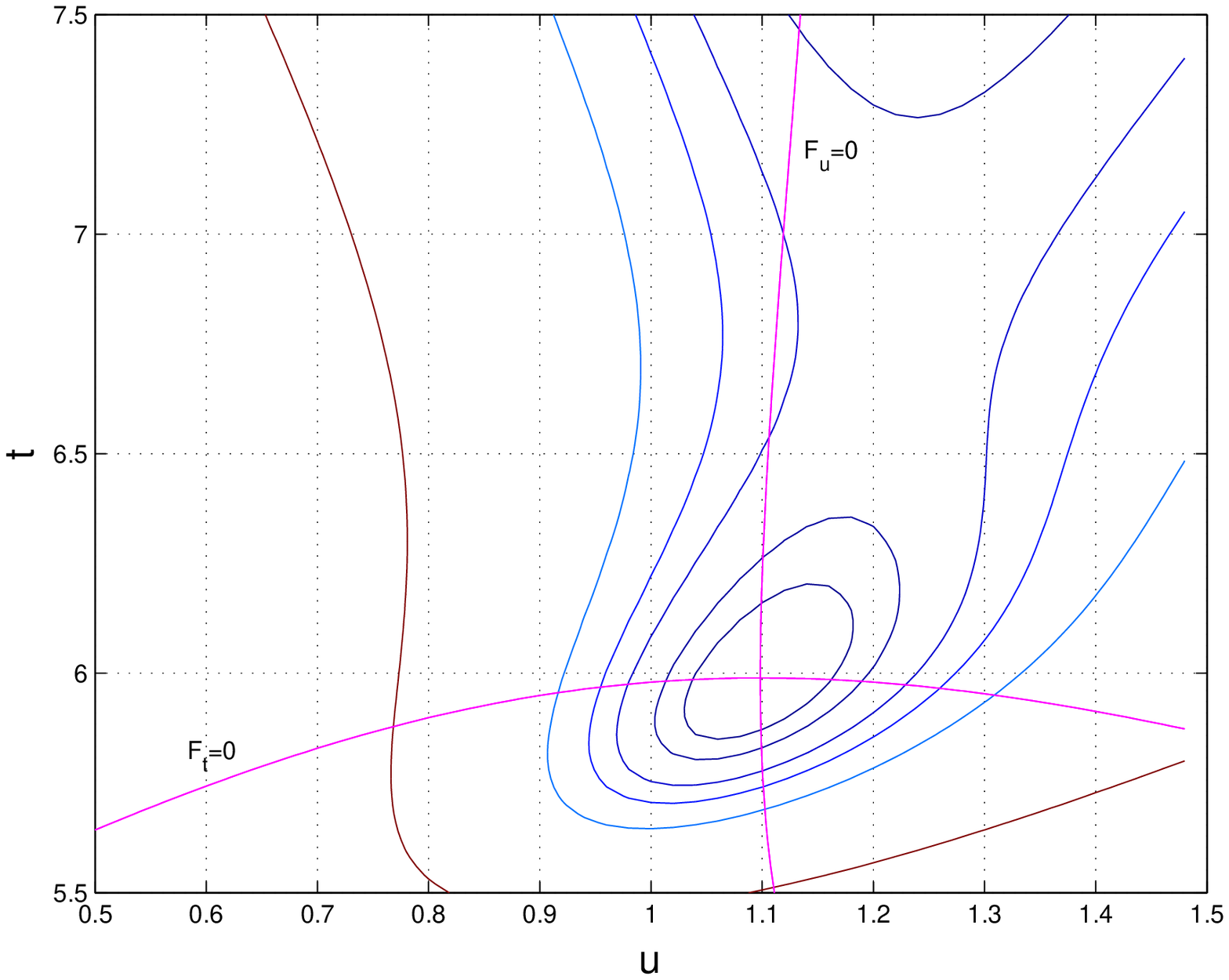}
\caption[fields]{\label{fig:ctuV0} Contour plot of the potential,
given by Eq.~(\ref{V0}), in the $t$, $u$ plane, for $m=3$,
$k=1200$, $\mu=-10$, $l=-30$. We have also plotted the conditions
$F_T=F_U=0$ in order to show the supersymmetric character of the
minimum. The imaginary parts of all fields have been set to zero.}
\includegraphics[width=10cm]{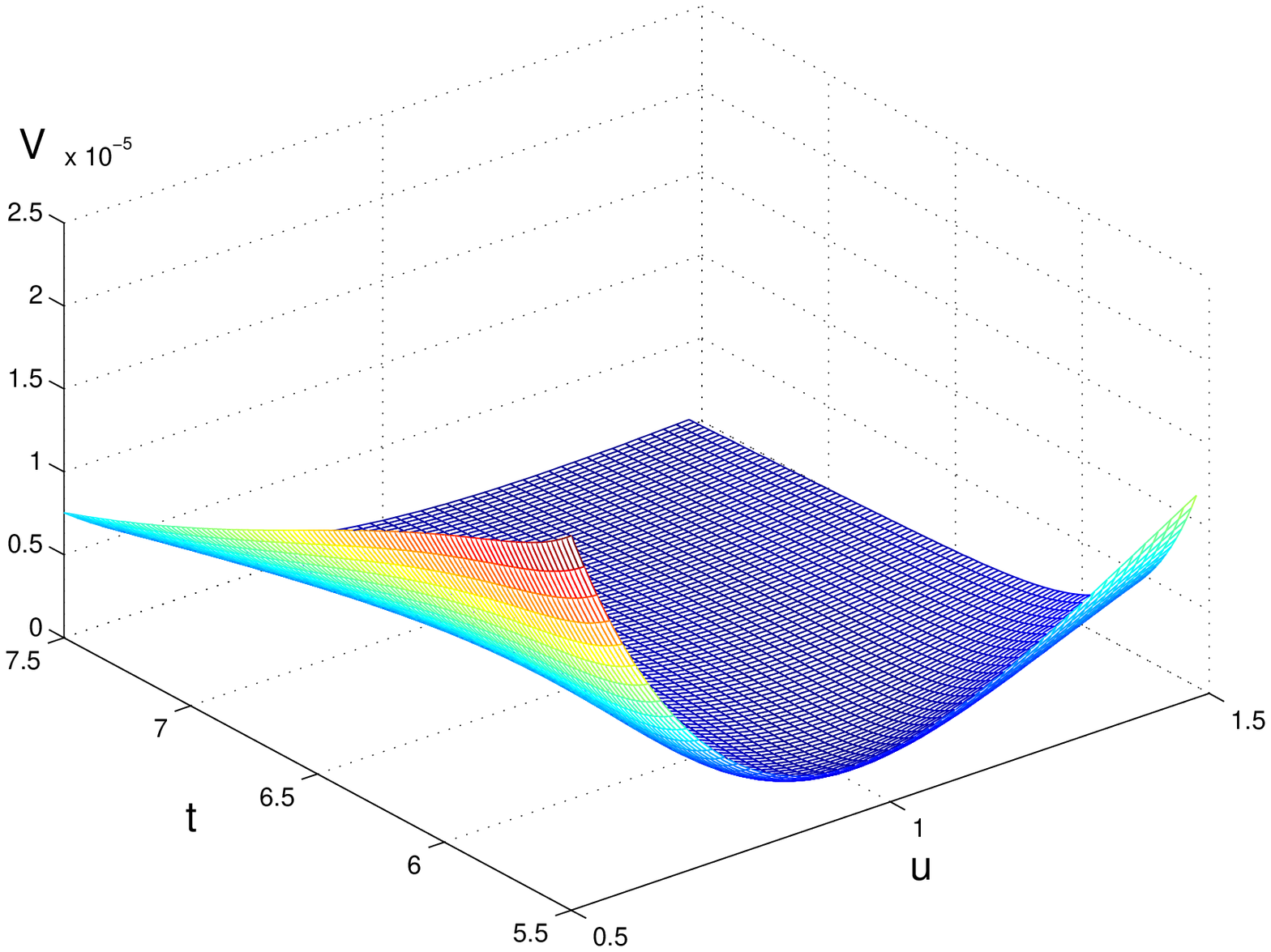}
\caption[fields]{\label{fig:3dtuV0} Plot of the potential, given
by Eq.~(\ref{V0}), as a function of $t$, $u$ for the same values
of parameters as in the previous figure. The minimum corresponds
to Eqs~(\ref{Tval},\ref{Uval},\ref{tuningTU}) being fulfilled.}
\end{center}
\end{figure*}
We confirm the previous statements by showing an explicit example
of a real supersymmetric minimum with vanishing cosmological
constant in Figures~\ref{fig:ctuV0},\ref{fig:3dtuV0}. The choice
of parameters is such that Eq.~(\ref{tuningTU}) is satisfied, and
the values of the fields correspond to $t=\ln(\k)=5.99$,
$u=\ln(l/\mu)=1.1$, in agreement with Eqs~(\ref{Tval},\ref{Uval}).
In Figure~\ref{fig:ctuV0}, the conditions $F_T=F_U=0$ are also
plotted to show that the minimum is indeed supersymmetric. Here we
are considering $b=8$, i.e. $I=1$, and we have checked that
corrections of order $\epsilon^2$ and higher do not affect the
existence and position of the minimum.

\vspace{0.4cm}

Let us now drop the condition of vanishing cosmological constant
and focus on real supersymmetric minima, that is, $\t =\n =0$. The
relevant F-terms have been given in Eqs.~(\ref{FTr}) and
(\ref{FUr}). It can be shown that solutions to the F-equations are
always minima for the model at hand. A very rough approximation to
the equation $\Re (F_U)=0$ leads to the solution
\begin{equation}
 u\simeq\ln\left|\frac{L}{M}\right|\; , \label{ususy}
\end{equation}
which coincides with the result~(\ref{Uval}) one obtains
for vanishing cosmological constant. Since we have set the phases
to zero, it exists only if $L$ and $M$ have the same sign.
Inserting this into the equation $\Re (F_T)=0$ we find $t$ is
determined by
\begin{equation}
 5+\k\left(2+\frac{7}{t}\right)e^{-t}+\frac{7M}{t}\left(1+\ln\left|\frac{L}{M}\right|\right)=0\; .
 \label{teq1}
\end{equation}
For sufficiently small $M$ this can be solved for negative $\k$
leading roughly to $t\sim t_{\rm max}$, where $t_{\rm max}$ was
defined in Eq.~(\ref{tmax}). Hence, we expect supersymmetric
minima for $\k <0$ and $L$, $M$ having the same sign. Further,
since $t_{\rm max}\sim \ln |\k |$, we typically need to have $|\k
|\gg 1$ in order to satisfy the constraint $b\e\ll 1$.
\begin{figure*}[!htb]
\begin{center}
\includegraphics[width=10cm]{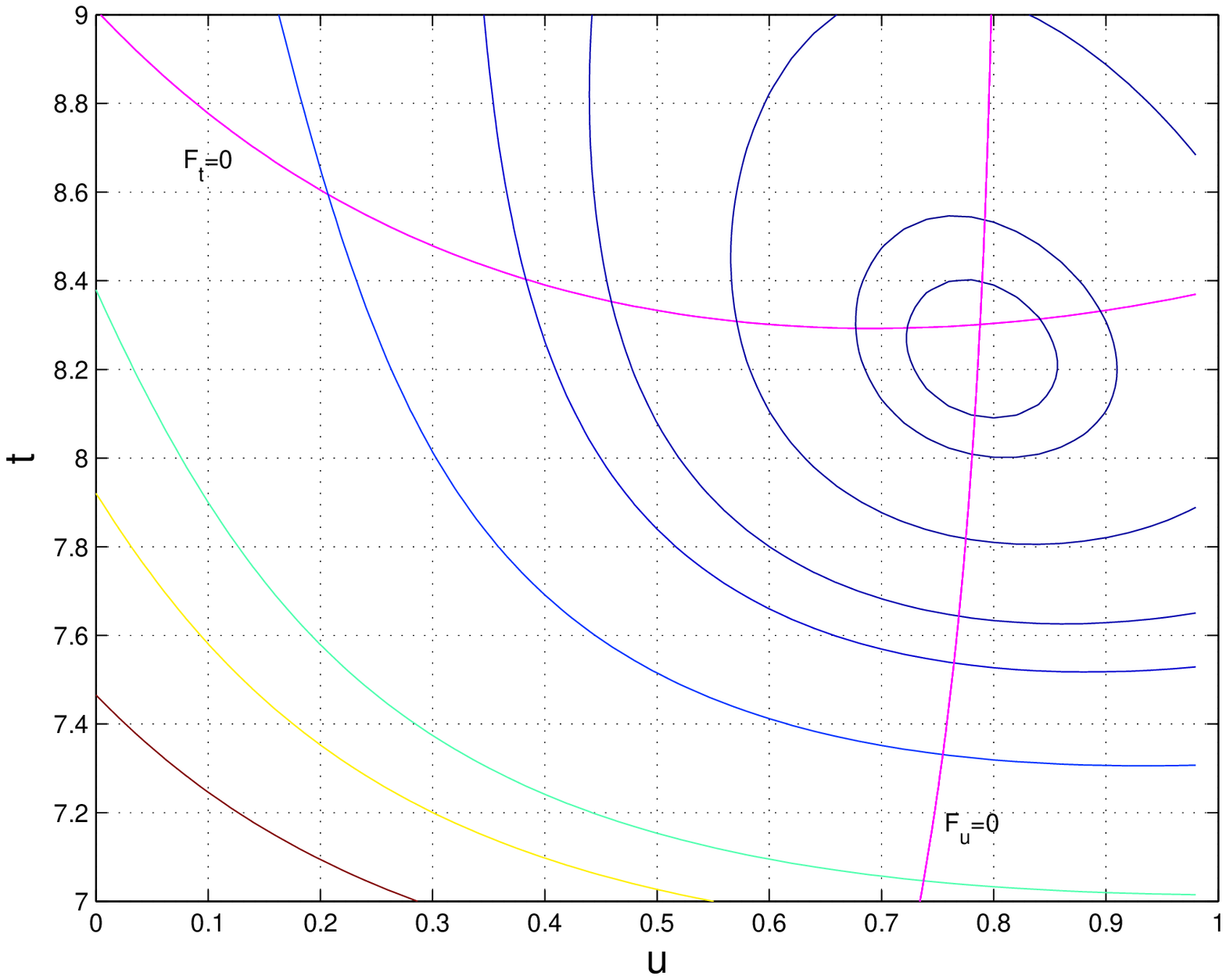}
\caption[fields]{\label{fig:ctususy} Contour plot of the
potential, given by Eq.~(\ref{V0}), in the ($t$, $u$) plane, for
$m=-1$, $k=1000$, $\mu=3$, $l=6$. We also plot the conditions
$F_T=F_U=0$ to show the supersymmetric character of the minimum.
The imaginary parts of all fields have been set to zero.}
\end{center}
\end{figure*}
This kind of supersymmetric minima, with negative cosmological
constant, is shown in Figure~\ref{fig:ctususy} for values of the
parameters given by $m=-1$, $k=1000$, $\mu=3$, $l=6$ (i.e.
$L,M<0$). The minimum in the $u$ direction is quite close to the
value given by Eq.~(\ref{ususy}), whereas Eq.~(\ref{teq1})
constitutes a rough approximation for the corresponding value of
$t$ (given by $t=8.25$). Similar minima can be obtained, as
already explained, for $\k<0$ and the opposite choice of signs for
$L,M$. And, again, we have chosen $I=1$ in order to keep our
perturbative expansion well under control.

For $\k >0$, using the relations (\ref{Tval}) with $p=0$ and
(\ref{tuningTU}), which describe the supersymmetric minima with
vanishing cosmological constant, Eq.~(\ref{teq1}) is identically
satisfied, as it should be. Perturbing away from this special
point in parameter space one can still find solutions to
Eq.(\ref{teq}) that are characterized by $\k>0$ and $M<0$. They
are precisely the supersymmetric minima close to the ones with
vanishing cosmological constant whose existence we have inferred
above from general argument.

In summary, we expect three classes of supersymmetric minima,
depending on the signs of the various parameters but all with
$L/M>0$. For $\k <0$ we can have either sign of $M$ while for $\k
>0$ supersymmetric solutions only exist for $M<0$.

\vspace{0.4cm}

We now turn to the search of minima with broken supersymmetry.
Unfortunately, here we can not count on the F-equations being
fulfilled in order to look for minima, however, as it turns out, a
good starting point is to look for solutions to the condition
$F_T=0$ in order to find minima of the potential (\ref{V0}). As it
has already been pointed out, in order to keep the supergravity
approximation, Eq.~(\ref{cons1}), valid, as well as to suppress
higher order terms in the K\"ahler potential, Eq.~(\ref{cons2}),
we need $t$ to be large at the minimum (at least of order 10).
This translates into a large value for $\k$. At the same time we
need to keep the value of $u$ of order 1, and the number of
blow-up moduli below about 10, in order to comply with the
condition $b\e^2 \ll 1$. This also guarantees the smallness of the
mixing terms between $t$ and $u$ in the scalar potential, and
explains why the minimization along the $t$ direction is almost
unchanged with respect to the purely supersymmetric case.

Therefore minima with broken supersymmetry appear for large $\k$ (and $F_T
\sim 0$) and $l$,$\mu$ of the same order of magnitude. As for
their signs, we have found only minima for $l<0$, $\mu>0$ whereas
the opposite choice gives rise to a runaway potential for
increasing values of $u$.
\begin{figure*}[!htb]
\begin{center}
\includegraphics[width=10cm]{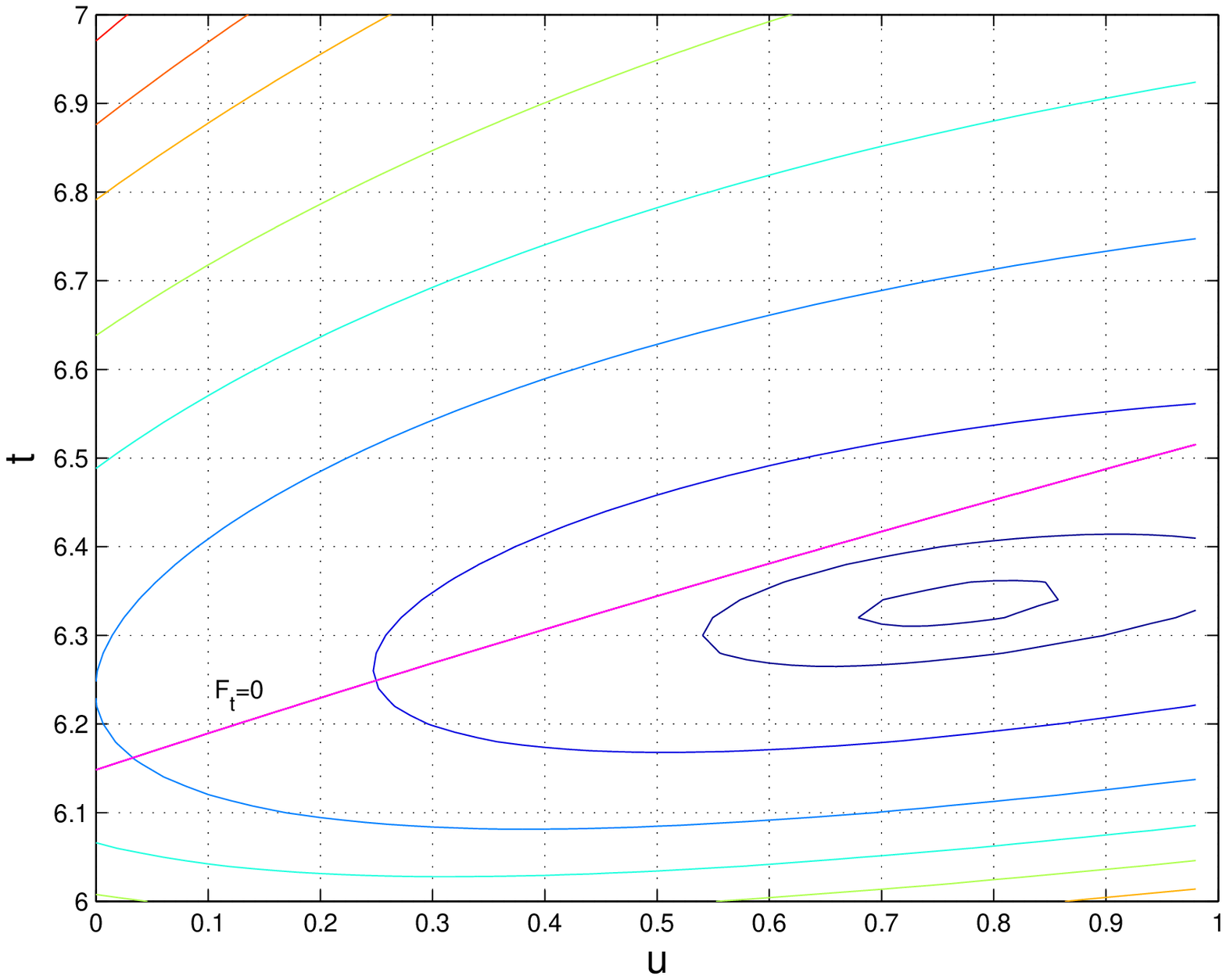}
\caption[fields]{\label{fig:ctunos} Contour plot of the potential,
given by Eq.~(\ref{V0}), in the ($t$, $u$) plane, for $m=-1$,
$k=1000$, $\mu=1$, $l=-1.5$. We have also added the condition
$F_T=0$. The imaginary parts of all fields have been set to zero,
where we have a minimum.}
\includegraphics[width=10cm]{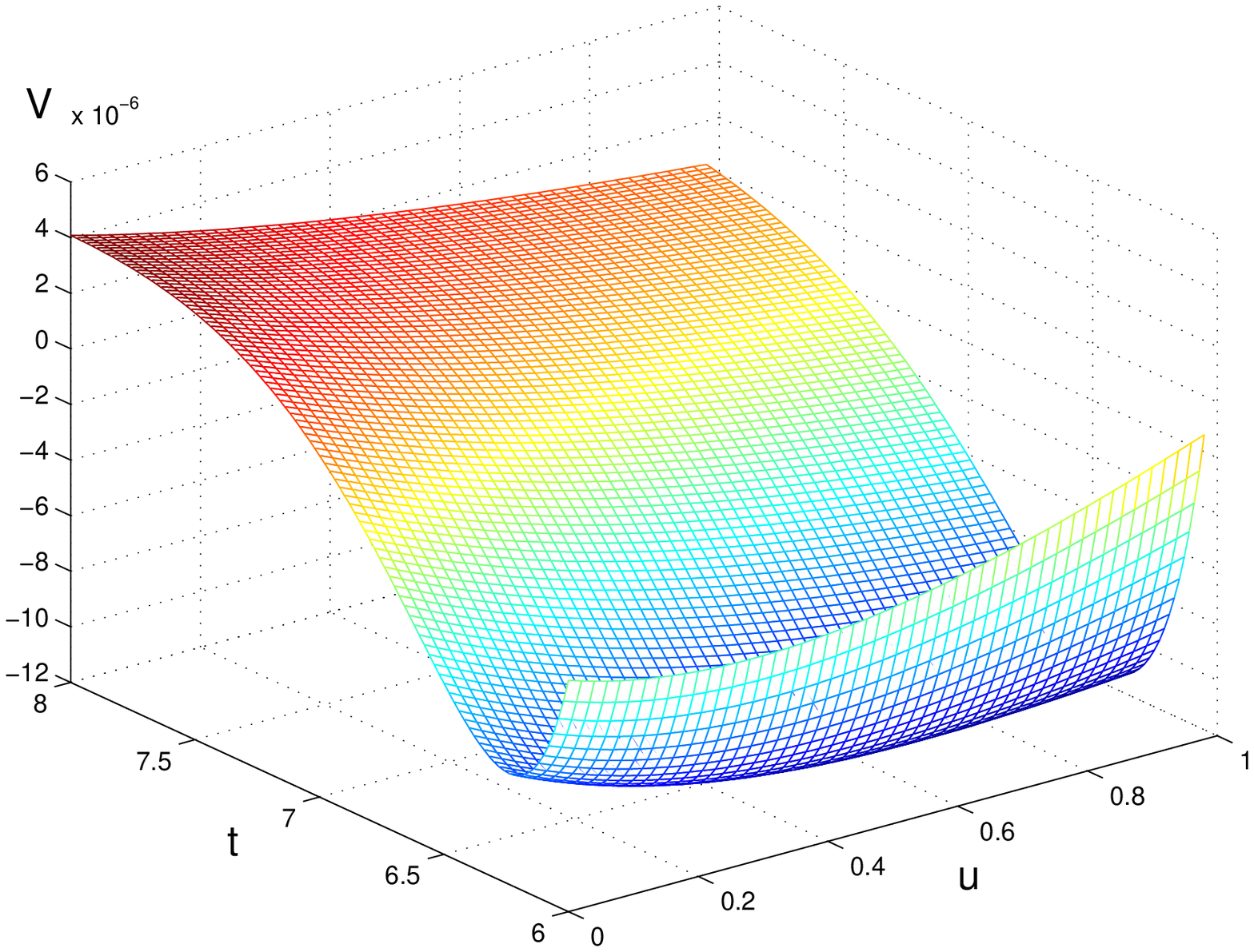}
\caption[fields]{\label{fig:3dtunos} Plot of the potential, given
by Eq.~(\ref{V0}), as a function of $t$, $u$ for the same values
of parameters as in the previous figure.}
\end{center}
\end{figure*}
A typical example of what has just been described is shown in
Figures~\ref{fig:ctunos},\ref{fig:3dtunos}, where we plot both the
contour and the shape of the potential given by Eq.~(\ref{V0}) for
$m=-1$, $k=1000$, $\mu=1$, $l=-1.5$ as a function of $t$ and $u$
(their imaginary parts having been set to zero). In
Figure~\ref{fig:ctunos} we have also plotted the constraint
$F_T=0$ in order to show how close it is to the minimum (whereas
$F_U=0$ lies well outside the plot). All the minima with broken
supersymmetry that we have found have a negative cosmological
constant.

Finally we would like to add a further comment on the stability of
our results. As already mentioned several times, we are imposing
the constraint $b\e^2 \ll1$ in order to guarantee that (as yet
unknown) higher order corrections to the K\"ahler potential will
not spoil the results presented here. This, in turn, means that we
have a tight restriction on the number of blow-up moduli (given by
$I$, with $b=8I$) allowed in our models. We have found that, in
order to achieve $u_{\rm min}\sim 1$, $I=1$ becomes almost the
only choice. We can still find minima with $0<u_{\rm min}< 1$ for
$I\leq 10$, however larger values of $I$ would result in the
minima shifting to negative values of $u$. There are, though,
plenty of examples in the literature of compact $G_2$ manifolds
with a small number of blow-ups, and while the precise K\"ahler
potential for those cases is not known, we expect it to resemble
the one used here. In that respect our results should be taken
from a purely phenomenological point of view. Once a complete
formula for the K\"ahler potential is known, it should be very
easy to incorporate it to our analysis, and it would be surprising
if the results differed substantially from those presented here.

\subsection{The general model}

We are now ready to analyze the general model defined by the
K\"ahler potential~(\ref{K}) and the superpotential~(\ref{W}),
which we write as
\begin{equation}
 W = \sum_{a=1}^7w_{(a)}+\sum_{i=1}^Iw_{(i)}\; ,
\end{equation}
where $w_{(a)}=w(m_a,k_a,T^a)$ and $w_{(i)}=w(\mu_i,l_i,U^i)$, and
the function $w$ has been defined in Eq.~(\ref{w}). The F-terms to
order $\e$ then take the form
\begin{eqnarray}
 F_a &=& w_{(a)}'-\frac{1}{2t_a}W \\
 F_i &=& w_{(i)}'+4f^iu_iW
\end{eqnarray}
with $f^i$ given in Eq.~(\ref{f}). For the scalar potential we find
\begin{eqnarray}
 V &=& \frac{1}{32\P_{c=1}^7t_c}\left\{\sum_{a=1}^7\left[t_a^2|w_{(a)}'|^2
      -t_a\Re (w_{(a)}'\bar{W})\right]+\sum_{i=1}^I\left[\frac{|w_{(i)}'|^2}{8f^i}
      -u_i\Re (w_{(i)}'\bar{W})\right]\right.\nonumber \\
   && \left.\qquad\qquad +2\sum_{a=1}^7\sum_{i=1}^Ip_{ia}t_au_i\Re (w_{(a)}'\bar{w}_{(i)}')
      +|W|^2\right\}\; .
\end{eqnarray}

\vspace{0.4cm}

We will not attempt a general classification of all minima of this
potential but rather present various classes of examples. We start
with the supersymmetric minima with vanishing cosmological
constant, characterized by the equations $w_{(a)}'=0$,
$w_{(i)}'=0$ and $W=0$. The field values are easily solved for and
are given by
\begin{eqnarray}
 T^a &=&\ln\left|\frac{k_a}{m_a}\right|+\p in_a\\
 U^i &=& \ln\left|\frac{l_i}{\m_i}\right|+\p in_i\; ,
\end{eqnarray}
where $n_a$ are even (odd) integers for $k_a/m_a$ positive
(negative) and, similarly, $n_i$ are even (odd) integers for
$l_i/\m_i$ positive (negative). Vanishing of the real and
imaginary parts of the superpotential implies
\begin{eqnarray}
 \sum_{a=1}^7m_a\left(1+\ln\left|\frac{k_a}{m_a}\right|\right)
 +\sum_{i=1}^I\m_i\left(1+\ln\left|\frac{l_i}{\m_i}\right|\right) &=& 0\label{tuninggen}\\
 \sum_{a=1}^7m_an_a+\sum_{i=1}^I\m_in_i &=& 0\; .
\end{eqnarray}
As in the analogous cases before, the first of these conditions is
the usual fine-tuning in parameter space required to set the
cosmological constant to zero, while the second has infinitely
many solutions for favourable signs of $k_a/m_a$ and $l_i/\m_i$
(or appropriate choices of the flux). We have hence found an
infinite number of supersymmetric minima with vanishing
cosmological constant that differ by integer shifts in the axion
directions. As before, all those minima correspond to vanishing
external flux, as Eq.~(\ref{extflux}) shows. In a neighbourhood of
the surface in parameter space defined by Eq.~(\ref{tuninggen})
there exist supersymmetric minima with negative cosmological
constant. We stress that the existence of these supersymmetric
minima does not depend on the form of the K\"ahler potential and
is, hence, not subject to the constraint~(\ref{cons2}) in moduli
space. Consequently, these minima exist for an arbitrary number
$I$ of blow-up moduli and, in particular, for the specific model
detailed in Table~\ref{tab1}.

\vspace{0.4cm}

In our discussion of the universal model with a single $T$ and $U$
modulus, we have encountered three classes of supersymmetric
minima characterized by the sign of the parameters. Above, we have
shown that the analog of one of these classes, namely the one that
includes supersymmetric minima with vanishing cosmological
constant, also exists in the general case. What about the other
two cases? We can construct examples for those cases by
generalizing the results for the universal case. We choose
specific parameters, such that
\begin{equation}
 k_a=k\; ,\qquad m_a=m\; ,\qquad l_i=\frac{7}{I}l\; ,\qquad\m_i=\frac{7}{I}\m\; ,\label{parid}
\end{equation}
for all $a$ and $i$ and universal parameters $k$, $m$, $l$ and $\m$. For these
parameters, we compute the F-terms at the universal point
\begin{equation}
 T^a=T\; ,\qquad U^i=U\; ,\label{fieldid}
\end{equation}
in field space. We find that $W=7\tilde{W}$ where
\begin{equation}
 \tilde{W} = w(k,m,T)+w(l,\m ,U)\; ,
\end{equation}
and
\begin{eqnarray}
 F_a &=& w'(k,m,T)-\frac{7}{2t}\tilde{W}\\
 F_i &=& \frac{7}{I}\left[w'(l,\m ,U)+\frac{bu}{t^2}\tilde{W}\right]\; ,
\end{eqnarray}
where $b=8I$. Comparison with the universal F-terms~(\ref{FT1}) and
(\ref{FU1}) then shows that every solution $(T,U)$ to the universal
F-equations for parameters $k$, $m$, $l$, $\m$ gives rise to a
solution of the general F-equations via the
identification~(\ref{parid}), (\ref{fieldid}). We still have to show
whether these solutions, if minima in the universal case, remain minima for
the general model. To do this, we evaluate the expressions~(\ref{V2})
and (\ref{V2m}) for the second derivative of the potential at
supersymmetric minima, using the general model defined by
Eqs.~(\ref{K}) and (\ref{W}) but specializing to universal
parameters~(\ref{parid}) and fields~(\ref{fieldid}). The result is
compared with that of an analogous calculation for the universal
$(T,U)$ model. This comparison shows that minima of the universal
$(T,U)$ model indeed remain minima of the general model at the
universal point if $t$ is sufficiently large and $\e$ small.  As
before, these minima still exist for non-universal parameters
sufficiently close to the universal choice~(\ref{parid}). We note that
the phenomenological parameter $b$ in the universal K\"ahler
potential~(\ref{KTU}) is identified as $b=8I$ under this
correspondence, where we recall that $I$ is the number of blow-up
moduli. In our analysis of the universal case we found that consistent
minima, stable under higher-order corrections to the K\"ahler
potential, exist for $b\lappeq 40$ ($b\lappeq 80$ if one tolerates a
small value of $u$). This translates into an upper bound of $I\lappeq
5$ ($I\lappeq 10$ allowing small values of $u$) for the maximal number
of blow-up moduli for which we can construct supersymmetric minima in
this way. In summary, we conclude that the supersymmetric minima with
negative cosmological constant can indeed be extended to complete
models with up to $5$ (or $10$) blow-up moduli.

\vspace{0.4cm}

Can the supersymmetry breaking minima with negative cosmological
constant we found in the universal case be generalized to the full
model?  Let us consider a minimum for the universal model with
parameters $m$, $k$, $\m$, $l$ and field values $T$ and $U$ and
analyze the general model at universal parameter values~(\ref{parid})
and universal field values~(\ref{fieldid}). It is straightforward to
show that
\begin{equation}
 \frac{\partial V}{\partial U^i} = 0\; ,
\end{equation}
using the fact that the $U$ derivative of the universal potential~(\ref{VWu}) vanishes
and identifying $b=8I$, as before. Unfortunately, the $T^a$ derivatives of $V$ do not
vanish exactly at the universal point, due to the non-trivial coupling
between the $T$ and $U$ moduli encoded in the coefficients $p_{ia}$. 
However, it can be shown that $\partial V/\partial T^a$ consists
of terms either of order $\e$ or suppressed by inverse powers of $t$.
Hence, for sufficiently large $t$ and small $\e$ the derivatives
$\partial V/\partial T^a$ are small and there will be an extremum
for field values close the universal choice. Moreover, under the same
conditions on $t$ and $\e$ these will still be minima. We conclude,
that non-supersymmetric minima with negative cosmological constant
can be generalized to the full model. As above, the constraint on the
parameter $b$, necessary for consistent minima of the universal model
to exist, translates into a bound of $I\lappeq 5$ ($I\lappeq 10$ allowing small
$u$ values) on the number $I$ of blow-up moduli for which we can obtain
such non-supersymmetric minima.   


\section{Conclusion}
\label{conclusion}

In this paper we have analyzed the vacuum structure of four-dimensional
$N=1$ supergravity theories originating from M-theory on $G_2$ spaces,
with a superpotential from flux and membrane instanton effects. We have
focused on $G_2$ spaces which are constructed by blowing up $G_2$ orbifolds.
These spaces have two different types of moduli, namely ``bulk'' moduli
$T^a$ associated with the underlying orbifold and ``blow-up'' moduli $U^i$
which measure the size and orientation of the blow-ups. 

We have been starting the analysis with a simple toy model consisting
of a single bulk modulus $T$, generalizing to seven bulk moduli $T^a$,
then including blow-up moduli, first in a universal model with a
single bulk and blow-up modulus $(T,U)$, and, finally, studying the
full model. Our main result is that minima with negative cosmological
constant can be found. They exist for both supersymmetry preserved as
well as broken and,
after appropriate tuning of parameters, supersymmetric minima with
vanishing cosmological constant exist for the general model, as well
as for most of the simpler toy models. In constructing consistent
minima we had to respect one technical constraints on the moduli
space. The typical ratio $\epsilon= u/t$ of the a bulk modulus $t$
and a blow-up modulus $u$ had to be smaller than one since the
K\"ahler potential has only been calculated to leading order in
$\e$. The supersymmetric minima with vanishing cosmological constant
are unaffected by this constraint since they only depend on the
superpotential. However, for all other minima it implies an
upper bound on the number $I$ of blow-up moduli. Moreover, we need
a large parameters $k_a$ multiplying the $T_a$ instanton
contributions in the superpotential in order to generate sufficiently
large values for the bulk moduli. We expect both restrictions
could be avoided if the exact K\"ahler potential was known.

\vspace{0.4cm}

Although we did obtain supersymmetry breaking minima we have not been
able to find any examples with a positive cosmological constant.  From
our experience a positive cosmological constant cannot be achieved in
the given setting, using a combination of flux and membrane
instantons. In analogy with the IIB construction of Ref.~\cite{Kachru:2003aw},
this can presumably be achieved by adding wrapped (anti) M5-branes to
our set-up. Work in this direction is in progress. 

\vspace{1cm} \noindent {\large\bf Acknowledgments} A.~L.~and
B.~d.~C.~ are supported by PPARC Advanced Fellowships. S.~M.~is
supported by a Royal Society postdoctoral fellowship.


\vskip 1cm
\appendix{\noindent\Large \bf Appendix}
\renewcommand{\theequation}{\Alph{section}.\arabic{equation}}
\setcounter{equation}{0}

\section{Full potential for a single bulk and blow-up modulus}
\label{appA}

We consider the case with a single bulk modulus $T$ and a single blow-up
modulus $U$, split into real and imaginary parts as
\begin{eqnarray}
 T&=&t+i\t\\
 U&=&u+i\n\; .
\end{eqnarray}
The model is defined by the K\"ahler potential and superpotential
\begin{eqnarray}
 K&=&-7\ln (T+\bar{T})+b\frac{(U+\bar{U})^2}{(T+\bar{T})^2} \label{KTU1}\\
 W&=&mT+ke^{-T}+\m U+le^{-U}\; ,\label{WTU1}
\end{eqnarray}
where $m$, $k$, $\m$, $l$ and b are real constants. We then find for the
F-terms
\begin{eqnarray}
 \Re (F_T) &=& -\frac{m}{2}\left[5+\k\left(2+\frac{7}{t}\right)\cos (\t )e^{-t}
             +\frac{7}{t}(Mu+L\cos (\n )e^{-u})\right] \label{RFT}\\
 \Im (F_T) &=& \frac{m}{2}\left[K\left(2+\frac{7}{t}\right)\sin (\t )e^{-t}
               -\frac{7}{t}(\t +M\n -L\sin (\n )e^{-u})\right] \\
 \Re (F_U) &\simeq& m\left[M-L\cos (\n )e^{-u}+\frac{bu}{t}(t+\k e^{-t}\cos (\t ))\right]\\
 \Im (F_U) &\simeq&\left[L\sin (\n )e^{-u}+\frac{bu}{t^2}(\t -\k\sin (\t )e^{-t}+M\n )\right]\; ,
 \label{IFU}
\end{eqnarray}
where
\begin{equation}
 \k = \frac{k}{m}\; ,\qquad L=\frac{l}{m}\; ,\qquad M=\frac{\m}{m}\; .
\end{equation}
The scalar potential is given by
\begin{eqnarray}
 V &=& \frac{C^2}{t^5}\left\{ 1+\frac{7\t^2}{t^2}+\k\left[ 5\cos (\t ) +(\cos (\t )
     -\t\sin (\t  ))\frac{7}{t}-\frac{14\t\sin (\t )}{t^2}\right] e^{-t}\right.
     \nonumber\\
     &&\qquad +L\left[\left(3\cos (\n ) -\frac{7\sin (\n )\t}{t}\right)\frac{v}{t}
     +\frac{7\cos (\n )}{t} -\frac{14\t\sin (\n )}{t^2}\right] e^{-u}\nonumber\\
     &&\qquad +K^2\left[1+\frac{7}{t}+\frac{7}{t^2}\right] e^{-2t}
       +7L^2\left[\frac{1}{2b}+\frac{u}{t^2}+\frac{1}{t^2}\right] e^{-2u}\label{VTU} \\
     &&\qquad +KL\cos (\t -\n )\left[\left(4+\frac{7}{t}\right)\frac{u}{t}+\frac{7}{t}
     +\frac{14}{t^2}\right] e^{-t-u} +M^2\left[\frac{7}{2b}+\frac{7\n^2}{t^2}\right]\nonumber\\
     &&\qquad +M\left[\frac{4u}{t}+\frac{14\t\n}{t^2}\right] +KM\left[\frac{3\cos (\t ) u}{t} -
       \frac{14\sin( \t )\n}{t^2}
       -\frac{7\sin (\t )\n}{t}+\frac{7\cos (\t ) u}{t^2}\right] e^{-t}\nonumber\\
     &&\left.\qquad +LM\left[\frac{7\cos (\n ) u^2}{t^2}+\frac{7\cos (\n ) u}{t^2}
        -\frac{7\cos (\n )}{b} -\frac{7\sin (\n )\n u}{t^2}-\frac{14\sin (\n )\n}{t^2}\right]
        e^{-u} \right\}\; .\nonumber
\end{eqnarray}


\end{document}